\documentclass{ws-procs9x6}
\usepackage{amsmath,amssymb}
\usepackage{epsfig,graphicx}
\usepackage{txfonts}
\usepackage[hang,bf,nooneline]{subfigure}
\usepackage[hang,bf,nooneline]{caption}
\usepackage{rotating}
\begin{document}

\title{THE DARK CONNECTION BETWEEN THE EGRET EXCESS OF DIFFUSE GALACTIC GAMMA RAYS,
THE CANIS MAJOR DWARF, THE MONOCEROS RING, THE INTEGRAL 511 keV ANNIHILATION LINE,
THE GAS FLARING AND THE GALACTIC ROTATION CURVE. }

\author{W. DE BOER}

\address{University of Karlsruhe \\
Postfach 6980, D-76131 Karlsruhe, Germany\\
$^*$E-mail: wim.de.boer@cern.ch}

\begin{abstract}
The  EGRET excess of diffuse Galactic gamma rays shows all the key features of dark
matter annihilation (DMA) for a WIMP mass in the range 50-100 GeV, especially the
distribution of the excess is compatible with a standard halo profile with some
additional ringlike substructures at 4 and 13 kpc from the Galactic centre. These
substructures coincide with the gravitational potential well expected from the ring
of dust at 4 kpc and the tidal stream of dark matter from the Canis Major satellite
galaxy at 13 kpc, as deduced from N-body simulations fitting to  the Monoceros ring
of stars. Strong independent support for this substructure is given by the gas
flaring in our Galaxy.
 The gamma rays from DMA are originating predominantly from the hadronization of
mono-energetic quarks, which should produce also a small, but known fraction of
protons and antiprotons. Bergstr\"om et al. an antiproton flux far above the
observed antiproton flux and they conclude that the DMA interpretation of the EGRET
excess can therefore be excluded. However, they used an isotropic propagation
model, i.e. the same diffusive propagation in the disk and the halo. It is shown
that  an anisotropic propagation model is consistent with the EGRET gamma ray
excess, the antiproton flux and the ratios of secondary/primary and unstable/stable
cosmic ray particles. Such an anisotropic propagation is supported by  the large
bulge/disk ratio of the positron annihilation line, as observed  by the INTEGRAL
satellite. In this case no need for new sources specific to the bulge are needed,
so the claimed evidence for strong DMA in the bulge from these observations is
strongly propagation model dependent.
 In the framework of Supersymmetry cross section predictions
for direct dark matter searches are presented taking into account the EGRET data,
the WMAP data and other electroweak constraints.
\end{abstract}


\bodymatter
\section{Introduction}

Cold Dark Matter (CDM) is well established by the high rotation speeds in galaxies
and clusters of galaxies. Recent cosmological measurements yield a dark matter (DM)
density of $22\pm2\%$ of the energy of the Universe \cite{wmap}. If this DM is
created thermally during the Big Bang the present relic density  is inversely
proportional to $\langle\sigma v\rangle$, the annihilation cross section $\sigma$
of DM particles, usually called WIMPS (Weakly Interacting Massive Particles), times
their relative velocity. The average is taken over these velocities. This inverse
proportionality is obvious, if one considers that a higher annihilation rate, given
by $\langle\sigma v\rangle n_\chi$, would have reduced the relic density before
freeze-out, i.e. the time, when the expansion rate of the Universe, given by the
Hubble constant, became equal to or larger than the annihilation rate. The relation
can be written as: \cite{jungman}
\begin{equation}
\Omega_\chi h^2=\frac{m_\chi n_\chi}{\rho_c} h^2\approx (\frac{3\cdot 10^{-27} cm^3
s^{-1}}{\langle\sigma v\rangle})\label{eq1}.
\end{equation}
For the present value of $\Omega h^2=0.105 \pm 0.008$, as measured by WMAP
\cite{wmap}, the thermally averaged total cross section at the freeze-out
temperature of $m_\chi/22$ must have been around $3\cdot 10^{-26} ~{\rm
cm^3s^{-1}}$. Note that $\langle\sigma v\rangle$ as calculated from Eq.~\ref{eq1}
is independent of the WIMP mass (except for logarithmic corrections) as can be
shown by  detailed calculations \cite{kolb}. If the s-wave annihilation is
dominant, as expected  in many supersymmetric models, then the annihilation cross
section is energy independent, i.e. the cross section given above is also valid for
the cold temperatures of the present universe \cite{susy}. Such a large cross
section will lead to a production rate of mono-energetic quarks\footnote{The quarks
are mono-energetic, since the kinetic energy of the cold dark matter particles is
expected to be negligible with the mass of the particles, so the energy of the
quarks equals the mass of the WIMP.} in our Galaxy, which is 40 orders of magnitude
above the rate produced at any accelerator. The fragmentation of these
mono-energetic quarks will lead to a large flux of neutrinos, photons, protons,
antiprotons, electrons and positrons in the Galaxy. From these, the protons and
electrons disappear in the sea of many matter particles, but the photons and
antimatter particles may be detectable above the background, generated by  cosmic
ray interactions with the gas in the Galaxy. Such searches for indirect Dark Matter
detection have been actively pursued. Recent  reviews and references to earlier
work can be found in Refs. \cite{bergstrom,sumner,Bertone:2004pz}.

Gamma rays have the advantage that they point back to the source and do not suffer
energy losses, so they are the ideal candidates to trace the dark matter density
and have a spectral shape characteristic for mono-energetic quarks. The charged
components interact with Galactic matter and are deflected by the Galactic magnetic
field, so they do not point back to the source. Therefore the charged particle
fluxes have large uncertainties from the propagation models, which determine how
many of the produced particles arrive at the detector. For gamma rays the
propagation is straightforward: only the ones pointing towards the detector will be
observed.

An excess of diffuse gamma rays compatible with dark matter annihilation (DMA) has
indeed been observed by the EGRET telescope on board of NASA's CGRO (Compton Gamma
Ray Observatory)\cite{hunter}. Below 1 GeV the CR interactions describe the data
perfectly well, but above 1 GeV the data are up to a factor two above the expected
background. The excess shows all the features of DMA annihilation for a WIMP mass
between 50 and 70 GeV \cite{us}. Masses up to 100 GeV are possible if one assumes
the cosmic ray energy spectrum to vary in the Galaxy. The excess was observed in all sky
directions, which would imply that DM is not dark anymore, but shining in gamma
rays \cite{us}. Of course, such an important observation needs to be scrutinized
heavily. Among the most important criticism was a paper by Bergstr\"om et
al.\cite{bergstrom1} claiming that the antiproton flux from DMA, using the DM
distribution from the EGRET excess, would be an order of magnitude higher than the
observed antiproton flux. However, they use an isotropic propagation model with the
{\it same} propagation in the disk and the halo. For the expected anisotropic
propagation models everything can be made consistent by having a faster propagation
perpendicular to the disk than in the disk, in which case the antiprotons from DMA
in the halo do not return to the disk. This demonstrates  the large
uncertainties from propagation models for indirect dark matter searches.

It is interesting to note that  anisotropic propagation models can easily
explain a large bulge/disk (B/D)
ratio of  the positron annihilation, as observed from the intensity of
the 511 keV annihilation line of thermalized positrons. These positrons are e.g. produced
in the $\beta$-decays of radioactive nuclei from supernova explosions.
A large B/D ratio is simply due to the different
geometries of the disk and the bulge: in the bulge the positrons and electrons can
annihilate and radiate before escaping to the halo, in the disk they enter the halo
much faster  without having time to radiate and/or annihilate.  In contrast, in isotropic
propagation models one can only explain the large bulge/disk ratio by a new source for
positrons for the bulge alone. Anisotropic propagation models have enough freedom
to explain the results without new sources.  So the claimed evidence for DMA from
these observations\cite{boehm}
 are strongly propagation model dependent.

The paper is organized as follows: in  section \ref{egret} we summarize the DMA
interpretation of the EGRET excess.  In section \ref{prop1} we discuss the problems
with the isotropic propagation models. An anisotropic propagation model, which
simultaneously describes the antiproton flux, the EGRET excess and the observations
concerning primary and secondary cosmic rays and cosmic clocks is discussed in
section \ref{prop2}, while in section \ref{integral}  the
consequences for the INTEGRAL excess of the 511 keV line  in this anisotropic propagation
model are discussed. In section \ref{direct} the constraints from
direct dark matter searches are discussed. Section \ref{conclusion} summarizes the
results.

\section{The DMA Interpretation of the EGRET Excess of diffuse Galactic Gamma
Rays}\label{egret}

\begin{figure}
\begin{center}
 \includegraphics [width=0.42\textwidth,clip]{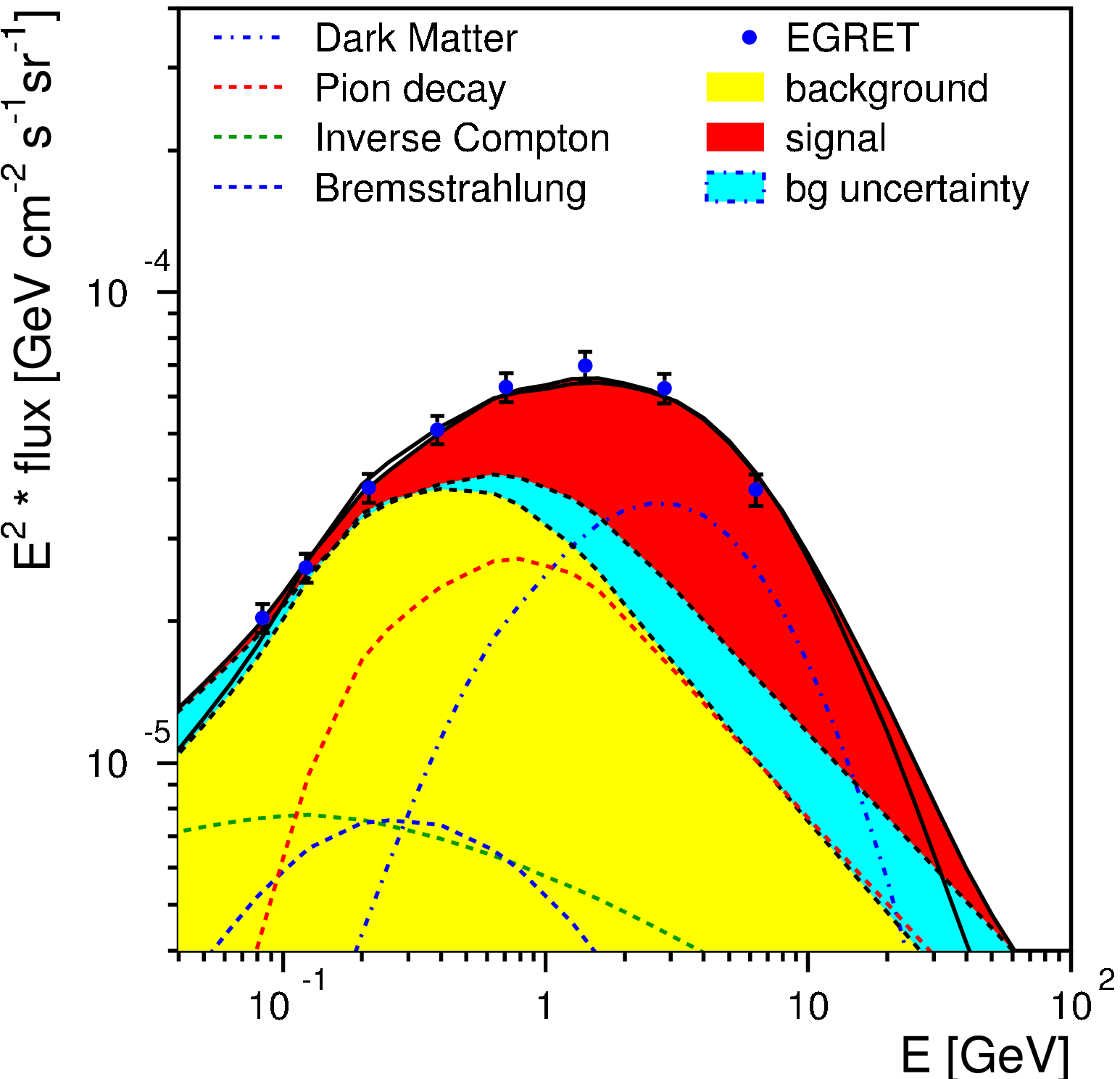}
 \includegraphics [width=0.49\textwidth,clip]{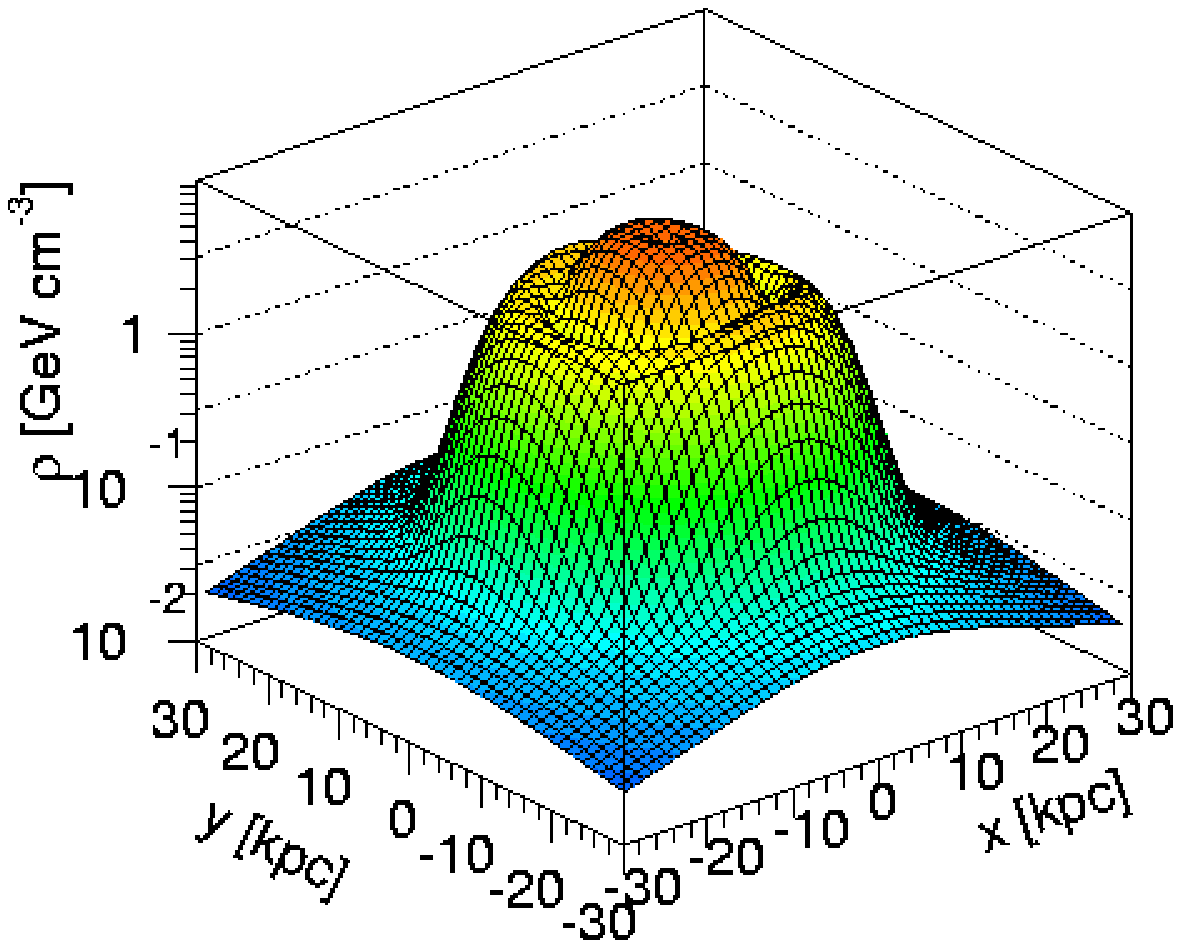}
\caption[]{Left: Fit of the shapes of background and
    DMA signal to the EGRET data in the direction of the Galactic centre
     The dark shaded (red)
    area indicates the signal contribution from DMA for a 60 GeV WIMP mass using the
    shape from high energy electron-positron annihilation experiments, while the
    area below it represents the background with the various indicated contributions.
    The (blue) area between the dotted lines just below the shaded area is the estimated
    uncertainty in the background, which is dominated by solar modulation.
    The normalization of both the background and the DMA contribution have been
    left free in the fit of the known shapes.
    Right: Parametrization of the dark matter density profile as determined from the distribution
    of the EGRET excess in the sky. Both pictures are taken from Ref. \cite{us}.}
 \label{fig1}
\end{center}
\end{figure}

 It is well known that the EGRET satellite data on diffuse gamma
rays shows an excess above 1 GeV in comparison with the expectations from CR
interactions \cite{hunter}. Below 1 GeV the CR interactions describe the data
perfectly well. The excess shows all the features of DMA annihilation for a WIMP
mass between 50 and 70 GeV \cite{us}. Especially, the two basic constraints
expected from any indirect DMA signal are fulfilled: \begin{itemize} \item the
excess should have the same spectral shape in all sky directions. \item the excess
should be observable in a large fraction of the sky with an intensity distribution
corresponding to the gravitational potential of our Galaxy. The latter means that
one should be able to relate the distribution of the excess to the rotation curve.
\end{itemize}
The analysis of the EGRET data is simplified by the fact that the spectral shapes
of the DMA contribution and the background from CR interactions with the gas of the
disk are well known from accelerator experiments: (i) the DMA signal should have
the gamma ray spectrum from the fragmentation of mono-energetic quarks, which has
been studied in great detail at LEP \cite{pdg}. (ii) the background in the energy
range of interest is dominated by CR protons hitting the hydrogen of the disk.
Therefore the dominant background spectral shape is known from fixed-target
experiments. Given that these shapes are known from the two best studied reactions
in accelerator experiments allows to fit these {\it known} shapes to the observed
gamma ray spectrum in a given sky direction and obtain from the fitted
normalization constants the contribution of both background and annihilation
signal. So in this case one does not need propagation models to estimate the
background, since the data itself calibrates the amount of background. A typical
spectrum is shown in Fig. \ref{fig1}, which clearly shows the rather distinct
shapes of DMA and background, so the two normalization constants are not strongly
correlated.  The shape of the DMA contribution shifts to the right for heavier WIMP
masses. The preferred WIMP mass is  between 50 and 70 GeV with a maximum allowed
value of 100 GeV.\cite{us}
\begin{figure}
\begin{center}
 \includegraphics [width=0.48\textwidth,clip]{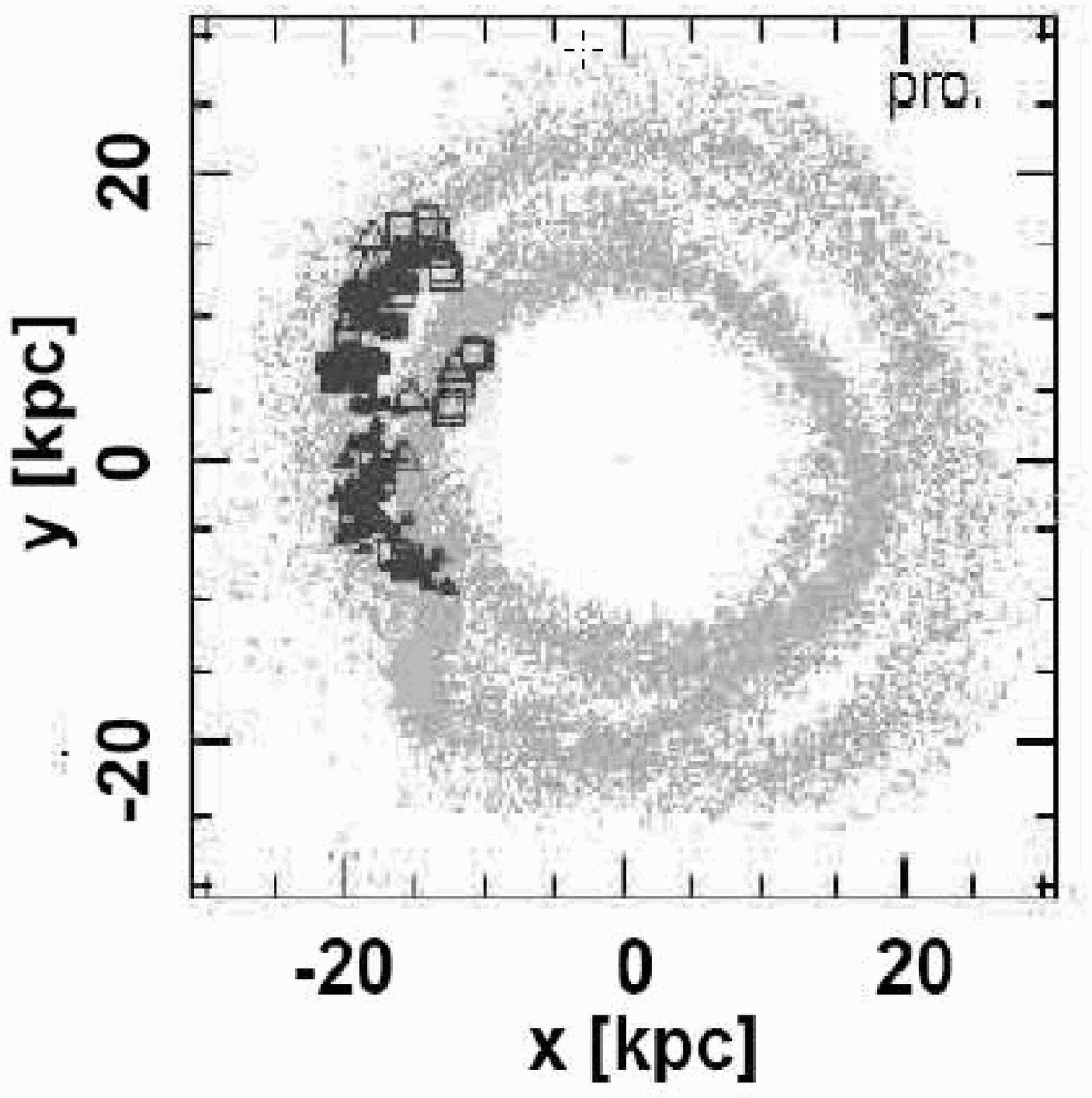}
 \includegraphics [width=0.45\textwidth,clip]{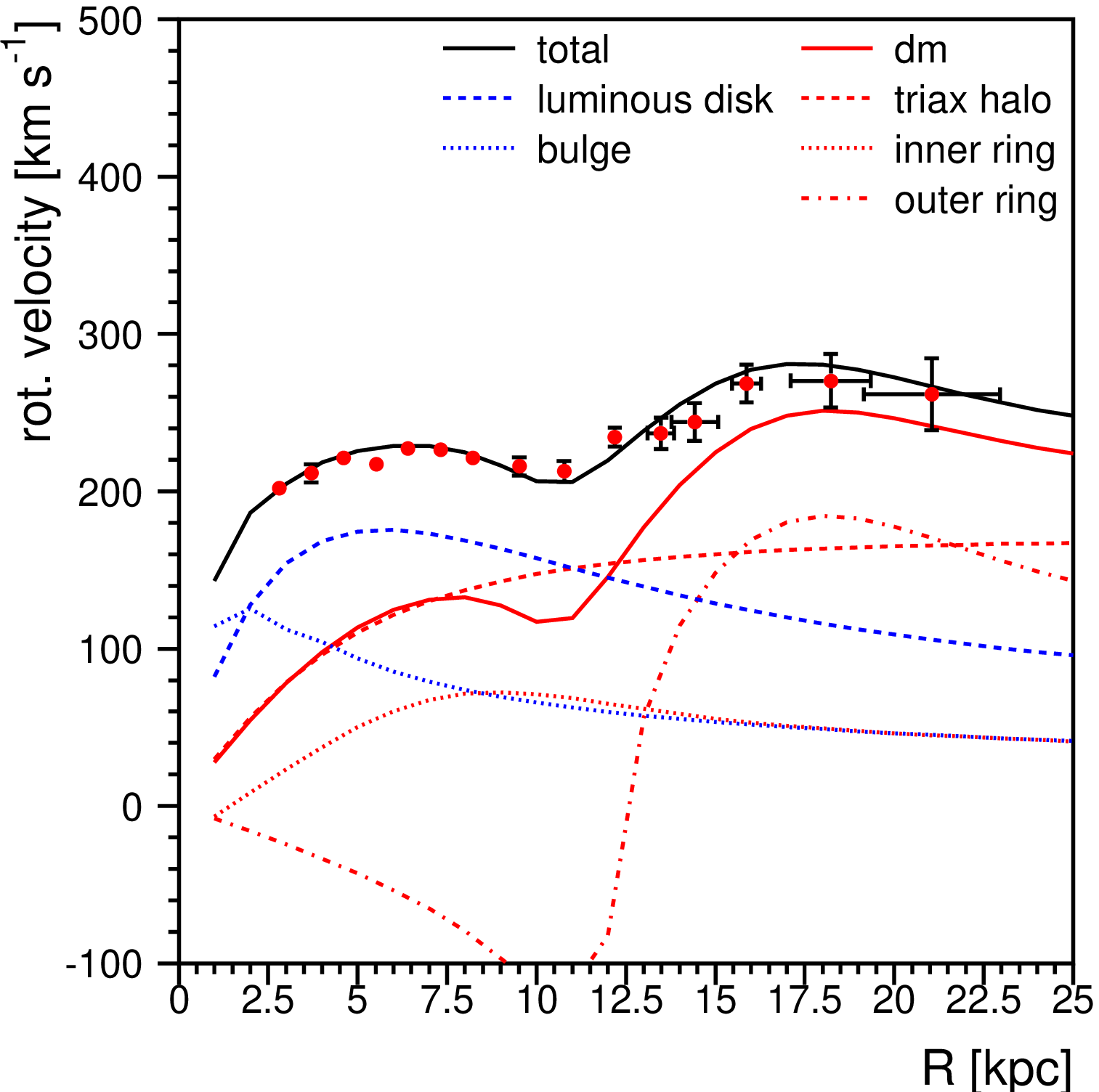}
 \caption[]{Left: results of an N-body simulation of the tidal disruption of the
 Canis Major dwarf Galaxy, whose orbit was fitted to the observed stars (symbols).
 The simulation predicts a ringlike structure of dark matter with a
 radius of 13 kpc. From Ref. \cite{penarrubia}.
 Right:The rotation curve with the contributions of the bulge, the disk, the triaxial dark matter
 halo and the two ringlike structures. The outer ring causes the peculiar change of slope in the
 rotation curve at about 11 kpc. From  Ref. \cite{us}.
    }
 \label{fig2}
\end{center}
\end{figure}
Repeating these fits over 180 independent sky directions showed that indeed: (i)
the {\it shape} of EGRET excess is consistent with a 60 GeV WIMP mass  in each sky
direction (ii) the {\it intensities} of the EGRET excess in various sky directions
are as expected from the gravitational potential, which could be proven by
reconstruc\-ting the rotation curve from the EGRET data after adding the known
distribution of visible matter to the DM halo \cite{us}. A parametrization of the
best fitted DM halo is shown in Fig. \ref{fig1} on the right hand side, which
clearly shows the ringlike structures at 4 and 13 kpc, as determined from the
enhanced intensity of the EGRET excess in these regions. The ring at 4 kpc (inner
ring) coincides with the ring of dust in this region. The dust is presumably kept
there because of a gravitational potential well, which is provided by the ring of
DM. The  ring at 13 kpc (outer ring) is thought to originate from the tidal
disruption of the Canis Major dwarf galaxy, which circles the Galaxy in an almost
circular orbit coplanar with the disk
\cite{ibata1,penarrubia,penarrubia1,ibata3,penarrubia2}. Three independent
observations confirm this picture of the ring originating from the tidal disruption
of a dwarf galaxy:

 (i) a ring of DM is expected in this region from the observed
ring of stars, called Monoceros ring, which was discovered first with SDSS data
\cite{newberg,yanny}. Follow-up observations \cite{ibata} found that this structure
surrounds the Galactic disk as a giant ring (observed over  100 degrees in
latitude) at Galactocentric distances from ~ 8 kpc to ~ 20 kpc. Tracing this
structure with 2MASS M giant stars, \cite{rocha-pinto} suggested that this
structure might be the result from the tidal disruption of a merging dwarf galaxy.
N-body simulations show indeed that the overdensity in Canis Major is indeed a
perfect progenitor for the Monoceros stream and they predict  a DM ring at 13 kpc
with a low ellipticity and almost coplanar with the disk, as shown in the left
panel of Fig. \ref{fig2}. The orientation of the major axis at an angle of 20
degrees with respect to the axis sun-Galactic centre and the ratio of minor to
major axis around 0.8 agrees with the EGRET ring parameters given in Ref.
\cite{us}. This correlation with the EGRET excess lends both support to the DMA
interpretation of the EGRET excess {\it and}  the interpretation that the Monoceros
stream originates
 from the tidal disruption of the
Canis Major (CM) satellite galaxy, thus rejecting the interpretation that the overdensity
of stars forming the Canis
Major dwarf  is a warp of the Galactic disk (see
discussions e.g. in Refs. \cite{momany,ibata2}). A further rejection from the hypothesis
that the CM overdensity is just due to the warp comes from the gas flaring discussed below,
which shows that the Monoceros stream is connected to an enormous amount of dark matter.

(ii) Such a massive ring structure influences the rotation curve in a peculiar way:
it decreases the rotation curve at radii inside the ring and increases it outside.
This is apparent from the change in direction of the gravitational force from the
ring on a tracer, since this force decreases the force from the galactic centre for
a tracer inside the ring, but increases it outside the ring. This is indeed
observed as shown in the right hand panel of Fig. \ref{fig2}, where the negative
contribution of the outer ring is clearly visible.

\begin{figure}[t]
\begin{center}
 \includegraphics [width=0.75\textwidth,clip]{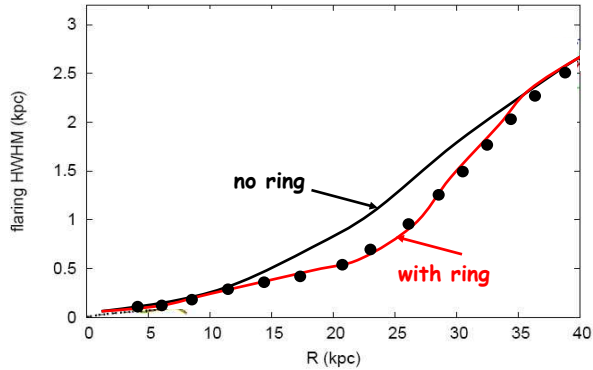}
 \caption[]{The half-width-half-maximum (HWHM) of the gas layer of the Galactic disk as function
 of the distance from the Galactic center. Clearly, the fit including a ring of dark matter
 above 10 kpc describes the data much better. Adapted from data in Ref. \cite{kalberla}.  }
 \label{fig3}
\end{center}
\end{figure}

(iii) A direct proof of the large amount of DM mass in the outer ring comes from a
recent analysis of the gas flaring in our Galaxy \cite{kalberla}. Using the new
data of the LAB survey of the 21 cm line in our Galaxy led to a precise measurement
of the gas layer thickness up to radii of 40 kpc. The increase of the half width of
the layer after a decrease to half its maximum value (HWHM)  as function of
distance is governed by the decrease in gravitational potential of the disk. The
outer ring increases the gravitational potential above 10 kpc, which is expected to
reduce the gas flaring. Only after taking the ring like structure into account the
reduced gas flaring in this region could be understood. The effect is shown in Fig.
\ref{fig3}. A fit averaged over all longitudes requires a DM ring with a mass of
$2.10^{10}$ solar masses, in rough agreement with the EGRET excess.

Clearly, these three independent astronomical observations need  a  ringlike DM
structure above 10 kpc, thus  providing independent evidence for the DMA
interpretation of the EGRET excess.

These independent observations
{\it cannot} be explained by alternative explanations of the EGRET excess, of which
the strongest one is provided by the "optimized" model\cite{optimized}. In this
case the cosmic ray spectrum of protons and electrons is not taken to be the
locally observed one, but modified to increase the gamma ray spectrum at high
energies. This requires a strong break in the injection spectrum of electrons and
protons in order not to change the gamma ray spectrum below 1 GeV, but only above 1
GeV. However, the gamma ray spectra
at intermediate and high latitudes are created by the {\it local} cosmic rays.
Still they show the excess. Furthermore, spectral shape differences are not
expected, since diffusion is fast compared with the energy loss time, so
diffusion equalizes the spectrum everywhere in agreement with the observation that
the gamma ray spectra in all directions can be described with the {\it same} CR
spectrum in all directions, i.e. the same spectral shape everywhere in the Galaxy.

Another  explanation is provided by tuning the efficiency of the EGRET spectrometer
to simulate DMA \cite{stecker}. However, this requires the efficiency already to
be modified around 1 GeV and reaching a change in efficiency of 80\% at 10 GeV in
clear disagreement with the calibration error in a photon beam before
launch\cite{egret_cal} and the residual uncertainties below 20\% during the flight
after correcting for time dependent effects\cite{egret_cal1}. Although their is
some uncertainty in the efficiency of the veto counter at higher energies because
of the backsplash from the calorimeter, this effect should not start at 1 GeV. Even
the authors agree that the considered effects are too small individually and it is
not clear that if one adds the errors all linearly that one gets effects up to
80\%.
And if these effects add up, it would be a remarkable coincidence  that the
excess corresponds exactly to the very specific sharply falling spectrum from the
fragmentation of mono-energetic quarks! An even more remarkable coincidence is that the distribution
of the excess in the sky follows
the gravitational potential, as proven by the  gas flaring and the rotation curve.
Therefore the excess is not as isotropic as suggested by these authors, if
one observes in finer sky bins, which reveals ringlike structures.
\section{The Antiproton Flux from DMA in an Isotropic Propagation Model}\label{prop1}
As mentioned in the introduction, a serious objection about the DMA interpretation
concerns the antiproton flux. However, this depends strongly on the propagation
model. Here we summarize the concepts used for the ``conventional'' propagation
model, discuss its priors and its alternatives.

The main features of our Galaxy are a barred central bulge with a diameter of a few
kpc and a large spiral disk of about 15 kpc. Most of the gas is distributed in the
disk, which extends to radii of  15-20 kpc, while the supernovae remnants (SNR)
peak at a distance of a few kpc from the centre. They are thought to be the source
of the cosmic rays (CRs) with energies up to $10^{15}$ eV. These CRs form a plasma
of ionized particles, in which the electric fields can be neglected by virtue of
the high conductivity and the magnetic fields  form  Alfv\'en waves, i.e. a
traveling oscillation of ions and the magnetic field. The ion mass density provides
the inertia and the magnetic field line tension provides the restoring force. The
wave propagates in the direction of the magnetic field with the Alfv\'en speed,
although waves exist at oblique incidence and smoothly change into  magnetosonic
waves when the propagation is perpendicular to the magnetic field. If the
wavelength of the Alfv\'en waves equals a multiple of the gyration radius of a CR,
resonant scattering occurs, which leads to a change in pitch angle of the CR (pitch
angle scattering) without changing its energy \cite{kulsrud}. Such a process leads
to a random walk of CRs, which can be described by a diffusion equation (see review
in Ref.\cite{strongrev}). If the B-field has no preferred direction, i.e. if the
turbulent small scale component is much stronger than the regular large scale
component, the waves propagate randomly in all directions and the diffusion of CRs
is isotropic. From the isotropy of CRs one usually assumes the propagation to be
isotropic.

Most primary nuclei show a power law spectrum falling with energy like $E^{-2.7}$.
This can be easily tuned by selecting the injection spectrum of the primary
particles accordingly. However, since the inelastic cross sections for secondary
particle production are usually not strongly energy dependent at higher energies,
this would lead to rather flat spectra for the secondary/primary ratios in contrast
to the observed B/C ratio, which shows a maximum at about 1 GeV/nucleon and
decreases as $E^{-0.6}$ towards higher energy. This can be accommodated by assuming
energetic particles diffuse faster out of the Galaxy, i.e. the diffusion constant
is proportional to $E^{0.6}$. This reduces the high energy part of the B/C spectra.
The decrease  at low energies can be accommodated by diffusive reacceleration,
which shifts the spectrum to higher energies \cite{strongrev}. Alternatively, one
can have a strong increase of the diffusion coefficient at low energies because of
the damping of the Alfv\'en waves \cite{ptuskin}, thus reducing the B/C ratio at
low energies as well.

Both stable and unstable nuclei are produced in supernovae (SNe) explosions with a
ratio given by their known production cross sections. From the decay time and the
remaining amount of the unstable nuclei one can reconstruct the  time of CRs needed
for their journey from the source to our local cavity, so they act as ``cosmic
clocks''.  Such measurements yield an average residence time of CRs in the Galaxy
of the order of $10^7$ yrs.  Since they travel with relativistic speeds the long
residence time requires that they cannot move rectilinear from the source to us or
to outer space, but the CRs must be scattering many times without loosing too much
energy, i.e. the diffusion must be effective. During their journey CRs may interact
with the gas in the Galaxy and produce secondary particles. This changes the ratio
of secondary/primary particles, like the B/C ratio. From the residence time and the
amount of secondaries one can estimate the grammage, i.e. the amount of matter
traversed  by a CR during its lifetime $t_{CR}$, which  is given by $\rho c
t_{CR}$, where $ct_{CR}$ is the path length for a particle traveling with the speed
of light $c$. It was found to be of the order of ${\rm 10 g/cm^{2}}$,
which corresponds to a density of about 0.2
${\rm atoms/cm^{3}}$.\cite{schlickeiser,berezinsky} This is significant lower than the
averaged density of the disk of
1 ${\rm atom/cm^{3}}$, which suggests that CRs travel a significant time in low density
regions, like the halo.

The most advanced program providing a numerical solution to the diffusion equation
is the publicly available GALPROP code \cite{galprop,galprop1,galprop2}.  The basic
parameters are the injection spectrum parameters, the diffusion coefficient,
the convection speed, the Alfv\'en speed and the size of the halo. The latter
determines the CR residence time inside the Galaxy, since as soon as they pass the
border, they are assumed to escape to outer space. By tuning these parameters to
the secondary/primary spectra and the unstable/stable spectra one obtains a
self-consistent propagation model of our Galaxy. The amount of secondary CR
particles and gamma rays are  described by this model by the cross sections of the
interactions of the primary and secondary CR with the gas of the disk using a
network with more than 2000 cross sections.
 This is one of the great triumphs of  GALPROP.

Remaining problems are connected with the large scale structure of the
propagation in the Galaxy, as sampled by gamma rays.
The problems are twofold.
First of all the  gamma rays in the GeV range, which are mainly produced by inelastic collisions
of CRs with the gas of the disk, show a too small radial gradient: large amounts of
gamma rays are produced at large longitudes, i.e. towards the Galactic anticentre.
This is unexpected, since the main sources of CRs are assumed to be the SNe
explosions, which are preferentially located at radii of a few kpc. In addition the
gas decreases at large radii and the gamma ray flux is proportional to the cosmic
ray density times the gas density. This problem can be remedied  by assuming much
more gas at large radii than determined from the molecular hydrogen ($H_2$) tracer,
which is the $\lambda=2.6$ mm line for the J=1 to J=0 transition of the carbon
monoxide (CO) molecule. This is a good tracer since CO gets excited by collisions
with $H_2$ molecules. Explaining additional gas at large radii requires  a strong
radial dependence of the ratio $X_{CO}=N(H_2)/N(CO)$, i.e.  $X_{CO}$ increases by
an order of magnitude from the inner Galaxy to the outer Galaxy \cite{strong_xco}.
An alternative explanation is provided by the Galactic wind models, in which case
the transport from  the disk to the halo is provided by the mass outflow from the
disk to the halo from the SNe explosions, which has a strong radial dependence
because of the strong radial dependence of the observed SNR \cite{breitschwerdt_gamma}.
This leads to strong
anisotropic propagation with the convection being strongest in the regions with the
highest cosmic ray pressure from the SNR, i.e. at distances between 4 and 10 kpc.
At smaller radii the gravitational potential of the bulge limits the outflow, while
at larger radii the CR pressure decreases \cite{breitschwerdt_gamma}. This reduces
the CR intensity in the disk and the corresponding gamma ray production  at the
position of the sources.

The second GALPROP problem is its failure to describe the EGRET excess of gamma rays, which
can be remedied by dark matter annihilation, as discussed in section \ref{egret}.
 If one attributes the EGRET excess to DMA one runs into the problem of a too large flux of
antiprotons, as discussed in detail by Bergstr\"om et al.\cite{bergstrom1}. We have
implemented the DMA as a source term into the publicly available GALPROP
code\footnote{The GALPROP code can be obtained from http://galprop.stanford.edu/.}
and find a similar result, as shown in the left panel of Fig. \ref{fig4} This is
not surprising, since GALPROP uses the same priors as the program used by
Bergstr\"om et al.\cite{bergstrom1}: (i) the propagation is dominated by diffuse
scattering, which is assumed to be isotropic, i.e. the same in the halo and the
disk (ii) the gas in the disk is smoothly distributed (iii) the influence of
observed regular magnetic fields can be neglected. These priors fulfill the basic
picture of the origin and propagation of cosmic rays discussed above. The main
reason for the large flux of antiprotons from DMA is {\it not} that DMA produces so
many antiprotons, but the fact that the residence time of charged particles is
required to be of the order of $10^7$ yrs, as determined from the cosmic clocks.
The last requirement can  be fulfilled in a model with isotropic diffusion only
with a large halo, so CRs do not escape, but they perform random walks in a large
volume. In this case there is no difference between primary particles produced by
SNe or primary particles produced by DMA, so {\it all} CRs are stored inside the
Galaxy. In this case DMA increases the averaged density of antiprotons by orders of
magnitude, so the flux of antiprotons becomes of the same order of magnitude as the
EGRET excess. Note that the production ratio of antiprotons/gammas from DMA is only
at the percent level, as is well known from accelerator experiments for the
fragmentation of mono-energetic quarks.  If one assumes that the propagation is
{\it not} isotropic, the picture completely changes: the DMA antiprotons may be
transported  quickly  to the halo by a combination of convection or fast diffusion
along the regular magnetic fields depicted in Fig. \ref{fig5}. Such an anisotropic
propagation model will be discussed in the next section.

\section{The Antiproton Flux from DMA in an Anisotropic Propagation Model}\label{prop2}

\begin{figure}[t]
\begin{center}
 \includegraphics [width=0.45\textwidth,clip]{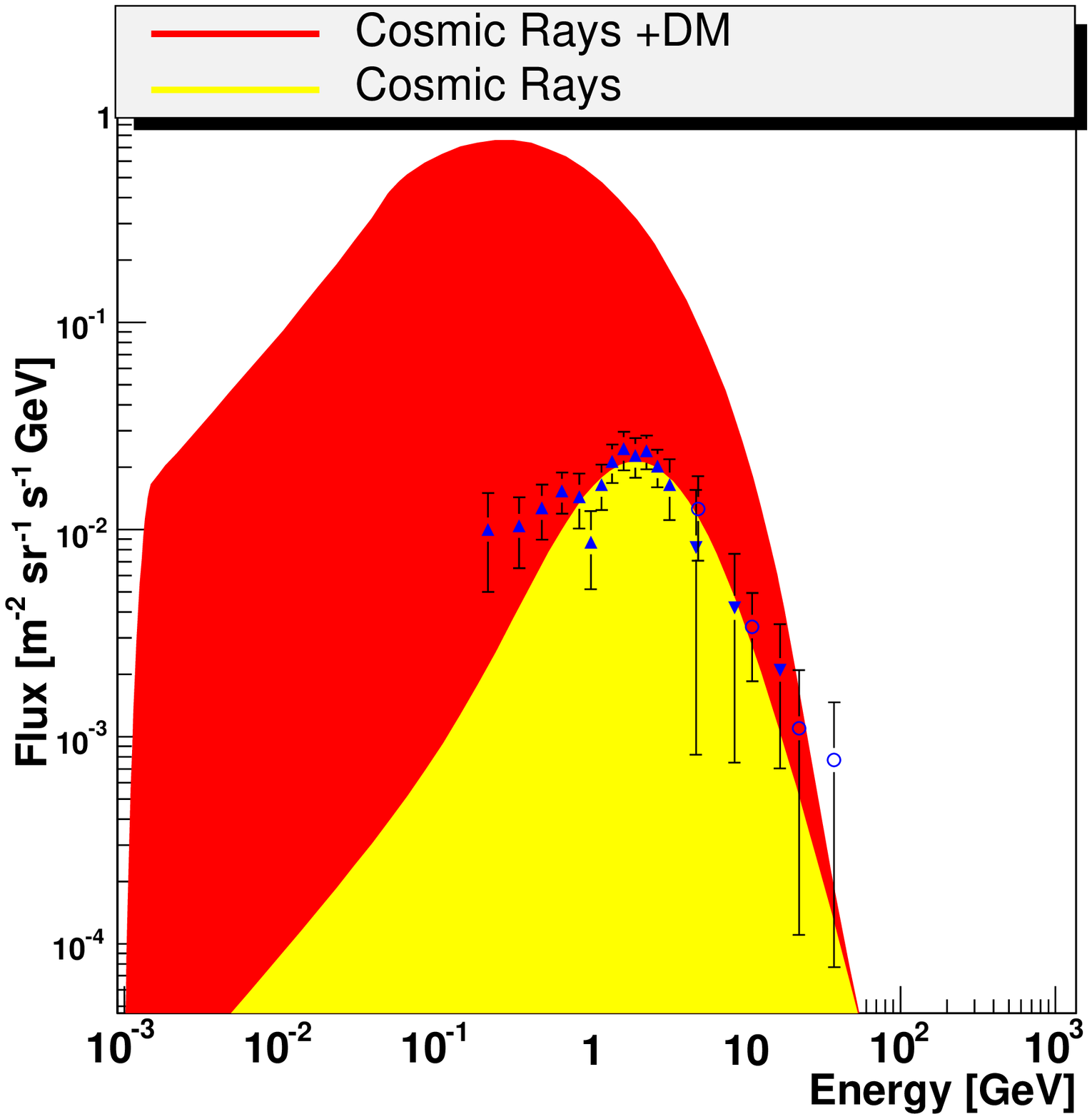}
  \includegraphics [width=0.45\textwidth,clip]{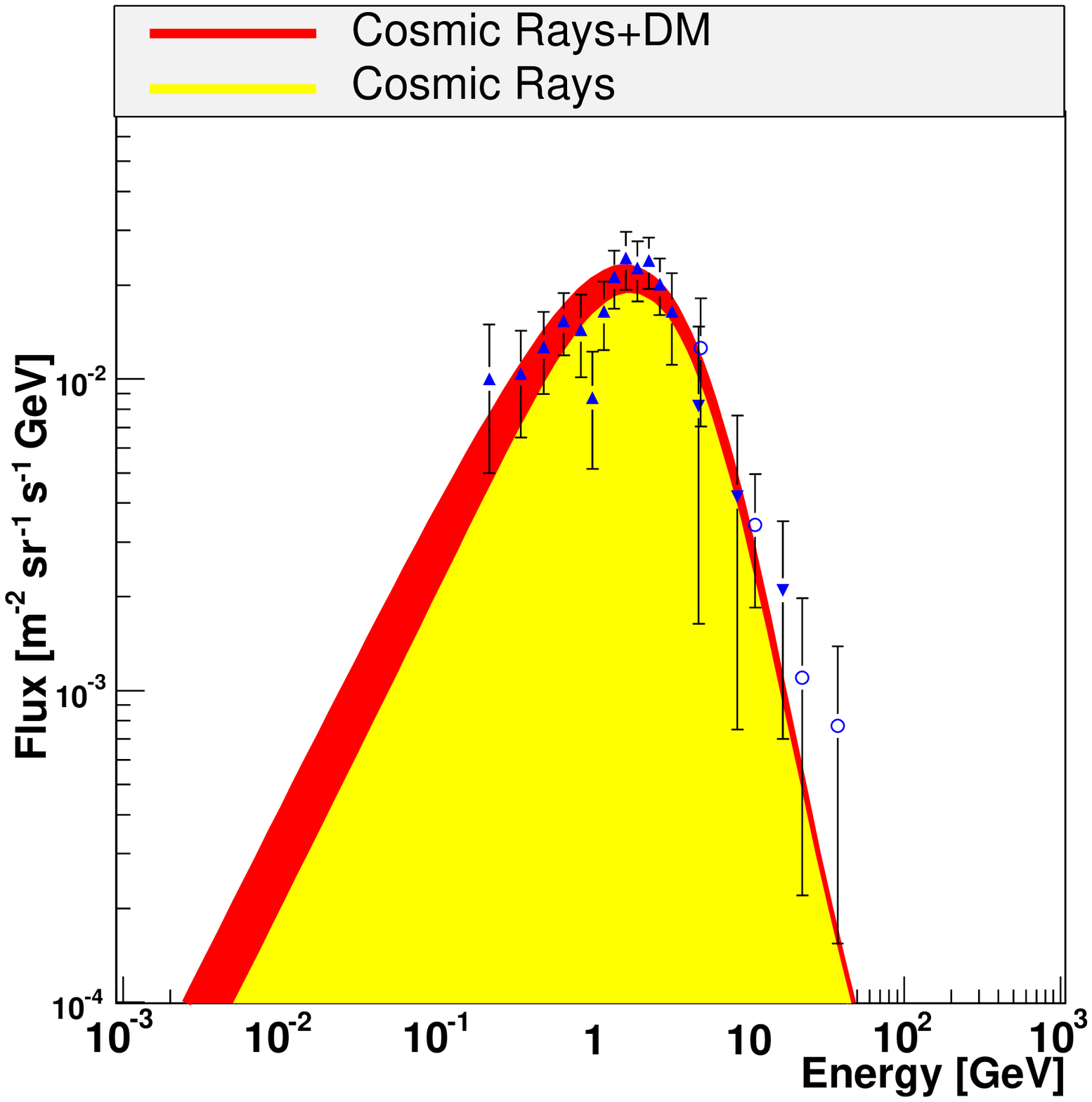}
 \caption[]{Comparison of the antiproton production including DMA
  in the conventional model with isotropic propagation (left)
 and a   model with anisotropic propagation (right).}
 \label{fig4}
\end{center}
\end{figure}

\begin{figure}[t]
\begin{center}
 \includegraphics [width=0.45\textwidth,clip]{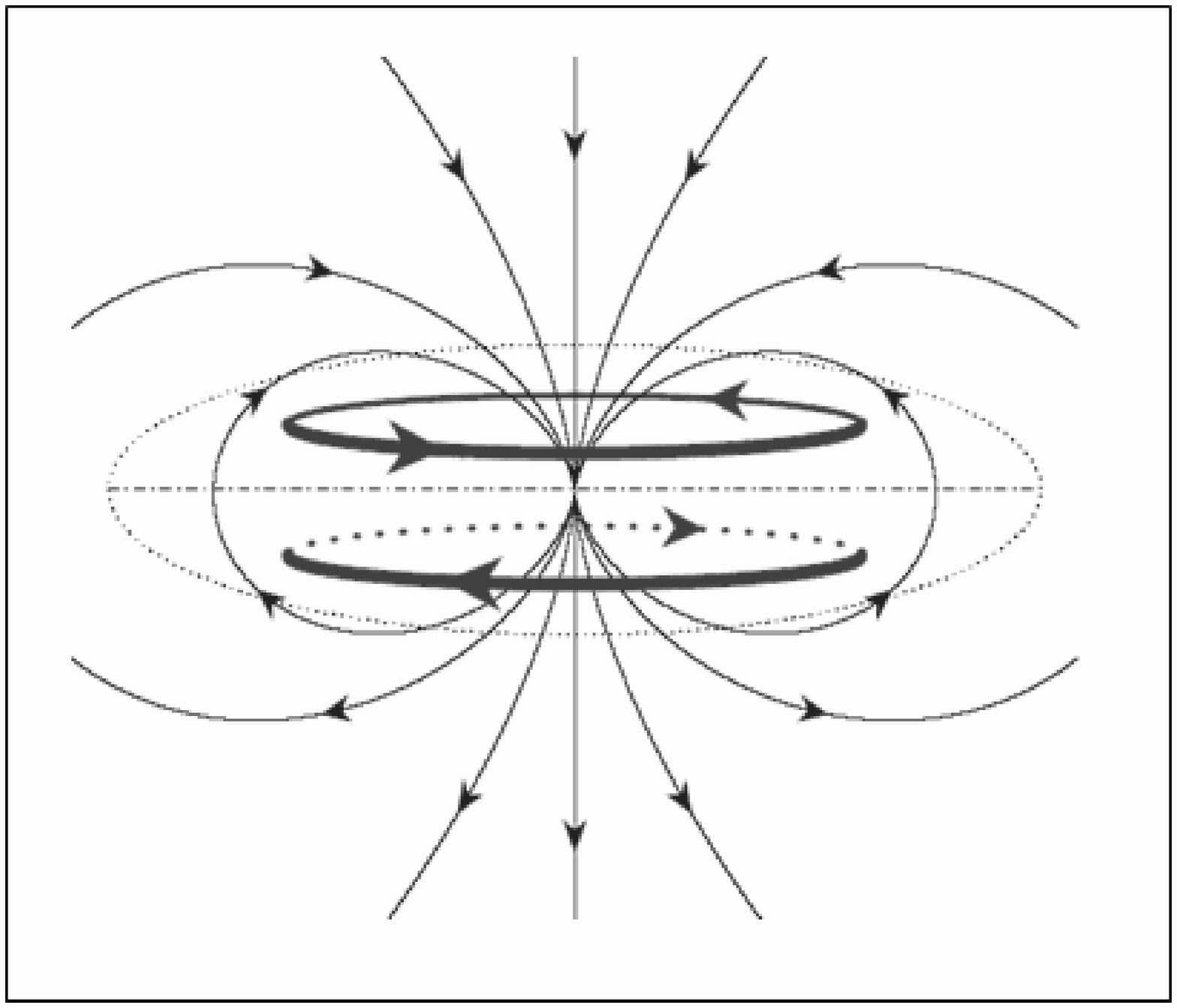}
 \includegraphics [width=0.45\textwidth,clip]{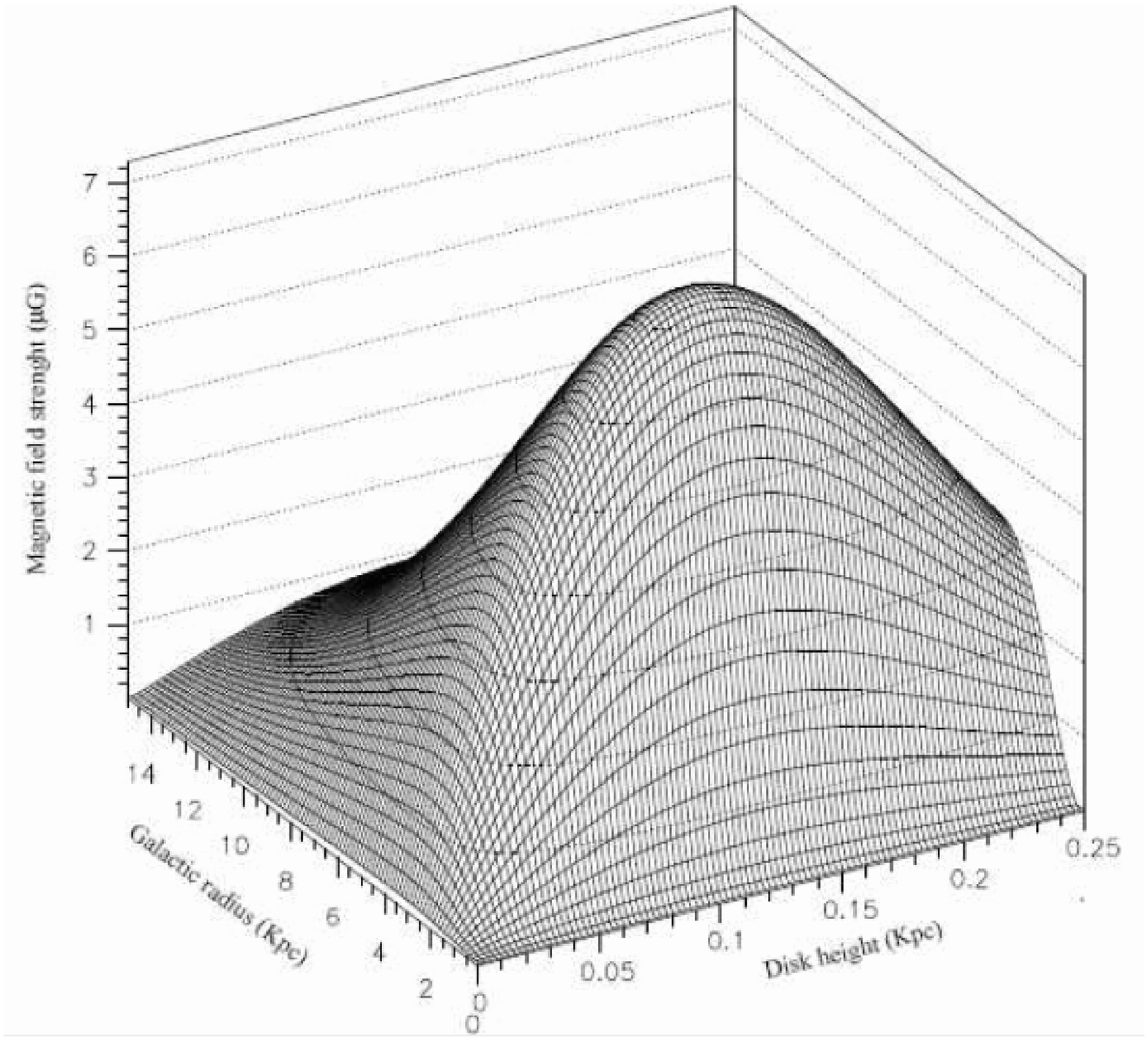}
 \caption[]{Left: a schematic picture of the magnetic fields in the Milky Way, consisting of a dipole (poloidal) field
 component (strongest in the center)  and
 a circular   (toroidal )  field component, which is strongest just above and below the disk
 at a distance of 150 pc (from Ref. \cite{han2}). Right: a parametrization of the toroidal component
 (from Ref. \cite{codino1}).}
 \label{fig5}
\end{center}
\end{figure}

The  propagation picture  with isotropic propagation, as discussed above, is based
on hydromagnetic wave theories, in which the turbulent small-scale components of
the magnetic field dominate over or are at the same order of magnitude as the
regular large scale components. However, the turbulence is expected to be different
in the halo and the disk, since the disk is not fully ionized in contrast to the
halo. This implies that the Alfv\`en waves are efficiently dissipated in the disk
by ion-neutral damping. Furthermore the most important contribution to the random
field in the disk is by turbulent mass motions, induced by supernova explosions and
other stellar mass loss activity, which leads to a different  wave spectrum in the
disk as compared to the halo, thus leading to different diffusion coefficients in
the halo and disk \cite{breitschwerdt_gamma}. In addition one expects the diffusion
along the regular magnetic fields to be an order of magnitude faster than the
diffusion perpendicular to the magnetic fields, even if the turbulent component is
of the same order of magnitude (see e.g. \cite{breitschwerdt_gamma} and references
therein). These analytical estimates of fast diffusion along the regular magnetic
field lines as compared to the transverse diffusion were confirmed by following the
trajectories of CRs in realistic magnetic fields with both a regular and turbulent
component. Given the toroidal field in the disk  \cite{blasi,codino1} found that
the diffusion along the azimuthal magnetic field is one to two orders of magnitude
faster than the transverse diffusion.  This means that CRs preferentially diffuse
along the toroidal fields just above or below the disk or into the halo via the
poloidal fields sketched in Fig. \ref{fig5}.\cite{han0,han1,han2} In such a picture
our local cavity is situated between two thick pancake-like structures of higher CR
density only 300 pc apart in the z-coordinate. This leads to a reduced grammage
compared with the CR density, which is maximal in the disk, a longer residence
time, a reduced radial gradient in the gamma ray flux and an isotropization of the
CR flux, all features difficult to explain simultaneously in an isotropic
propagation models with a source distribution located towards the centre.

Another possible effect concerning charged particles diffusing fast along regular
magnetic field components is related to the molecular clouds: the gas density in
the disk varies from $10^{-3}~ cm^{-3}$ in the warm ionized medium to
$10^2$-$10^3~cm^{-3}$ in clumps of cold gas with a size of a few pc. In the centre
of these clumps the density may be as high as $10^7~ cm^{-3}$ in dense molecular
clouds (MC), where star formation occurs. On average the gas density is $1~
cm^{-3}$ in the disk. Inside these MC magnetic fields far above the random
components have been observed (see Ref. \cite{heiles} for an excellent review). What is
more important, these fields seem to be correlated with the observed static
magnetic fields outside the MCs\cite{han}.  This can only be understood, if the MCs
remember the large scale magnetic fields in the interstellar medium, i.e. if during
the contraction flux freezing occurs. In this case the magnetic field lines from
the ISM will become highly concentrated near the MCs and CRs in the ISM, which
preferentially diffuse fast along  these field lines, will be reflected by the
higher density of the field lines near the MCs. So, as worked out by Chandran\cite{chandran}.
the MC act like magnetic mirrors for CRs, just like the
concentration of magnetic field lines near the poles from the earth trap the CRs in
the famous Van Allen radiation belts. The large distances (10-100 pc scale) between the MC
allows to trap particles up to the TeV scale, thus increasing the grammage and the
residence time, which  are now increased by the trapping time between the MC in the
disk, {\it not} by how often they pass from the halo to the disk, as is the case in
the isotropic propagation model. So the halo size is not a sensitive parameter
anymore and particles, once in the halo, will be preferentially transported away
from the disk by a combination of convection or fast diffusion along the regular
field lines in the halo. It should be noted that only a small fraction of the CRs
enter the MCs, if these act as magnetic mirrors. This could explain why  positrons
mainly annihilate in the gas {\it between} the MCs, not with electrons from
molecular hydrogen {\it inside } the MCs, as deduced from the annihilation line
shape \cite{jean,jean1}. Of course, one could argue that although the MCs make up
the largest mass fraction, they have the lowest filling factor. But if they  have
strong magnetic fields, the CRs are expected to be tunneled towards the MCs by the
high concentration of magnetic field lines.

How can one implement such a propagation model? Existing programs are not suitable.
E.g. the CR tracing programs calculating the trajectories in a magnetic field do
not calculate all secondary particles and the GALPROP program with isotropic
diffusion does not have a regular magnetic field in the propagation of charged
particles. However, the basic modifications needed are: the propagation follows
preferentially the regular magnetic field lines, which are toroidal in the disk and
poloidal in the halo. Such a propagation would require tuning the 3D version of
GALPROP. However, this takes an excessive amount of CPU time. Therefore
transporting the CRs to the halo by a combination of fast diffusion and convection
in the z-direction  in the 2D version is much more effective and leads to the same
effect: CRs  in the halo will hardly come back to the disk. But it should be kept
in mind that the parameters  are
 effective parameters in a simplified axisymmetric 2D version of the 3D reality,
 which has also preferred diffusion directions in the disk.

We have modified the publicly available source code of GALPROP by (i) allowing for
a diffusion tensor instead of a diffusion constant, thus modifying the diffusion
equation and the Crank-Nickelson coefficients accordingly; (ii) allowing an
inhomogeneous grid in order to have step sizes below 100 pc in the disk region and
large step sizes in the halo; (iii) implement the dark matter annihilation  as a
source term of stable primary particles, especially antiprotons, positrons and
gamma rays. The dark matter distribution was taken to be the one obtained from the
EGRET excess, as discussed above. The grammage and escape time were adjusted for
charged particles to account for the fact that secondary particles are now produced
largely locally, since particles produced far away from the solar system are likely
not to reach our local cavity. If in addition the trapping between molecular clouds
is effective, only a small fraction of CRs will penetrate the MCs and most of them
will be reflected. But the grammage in between the MCs will be enhanced by the
multiple passages from the mirroring, so one would expect it to increase the
grammage and residence time by the same factor. This turns out to be working, so
this was simply introduced in GALPROP as a constant  g, called grammage parameter,
multiplying the HI and HII gas densities. A grammage paramater of about 12 is
needed to describe the B/C ratio combined with the $^{10}Be/^9Be$ ratio.
Note that this grammage parameter
determines the local production of charged particle, so this grammage parameter is
not necessarily  the same as the grammage needed for the gamma ray production,
since the latter is determined by the large scale gas densities. Given the large
fluctuations in and strong radial dependence of the gas densities the difference
can be large. Also the CR density is expected to have a  radial dependence because
of the convection having a radial dependence, as discussed before. The CR density
can also have strong variations from the fact that Alfv\'en waves can be
damped in high density gas regions, thus leading to strong variations in the diffusion
coefficient and corresponding CR density variations.

\begin{figure}[t]
\begin{center}
 \includegraphics [width=0.45\textwidth,clip]{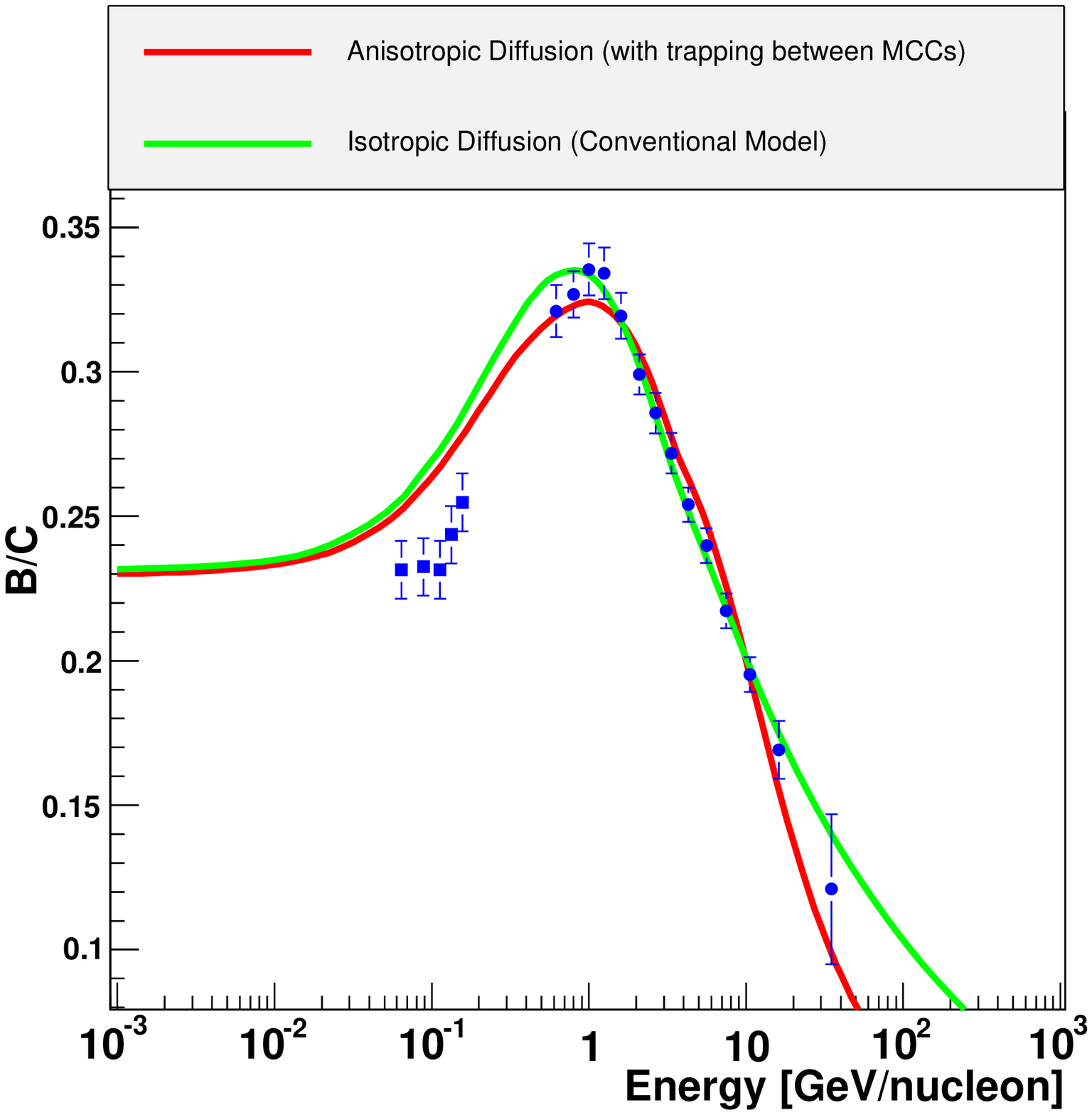}
  \includegraphics [width=0.45\textwidth,clip]{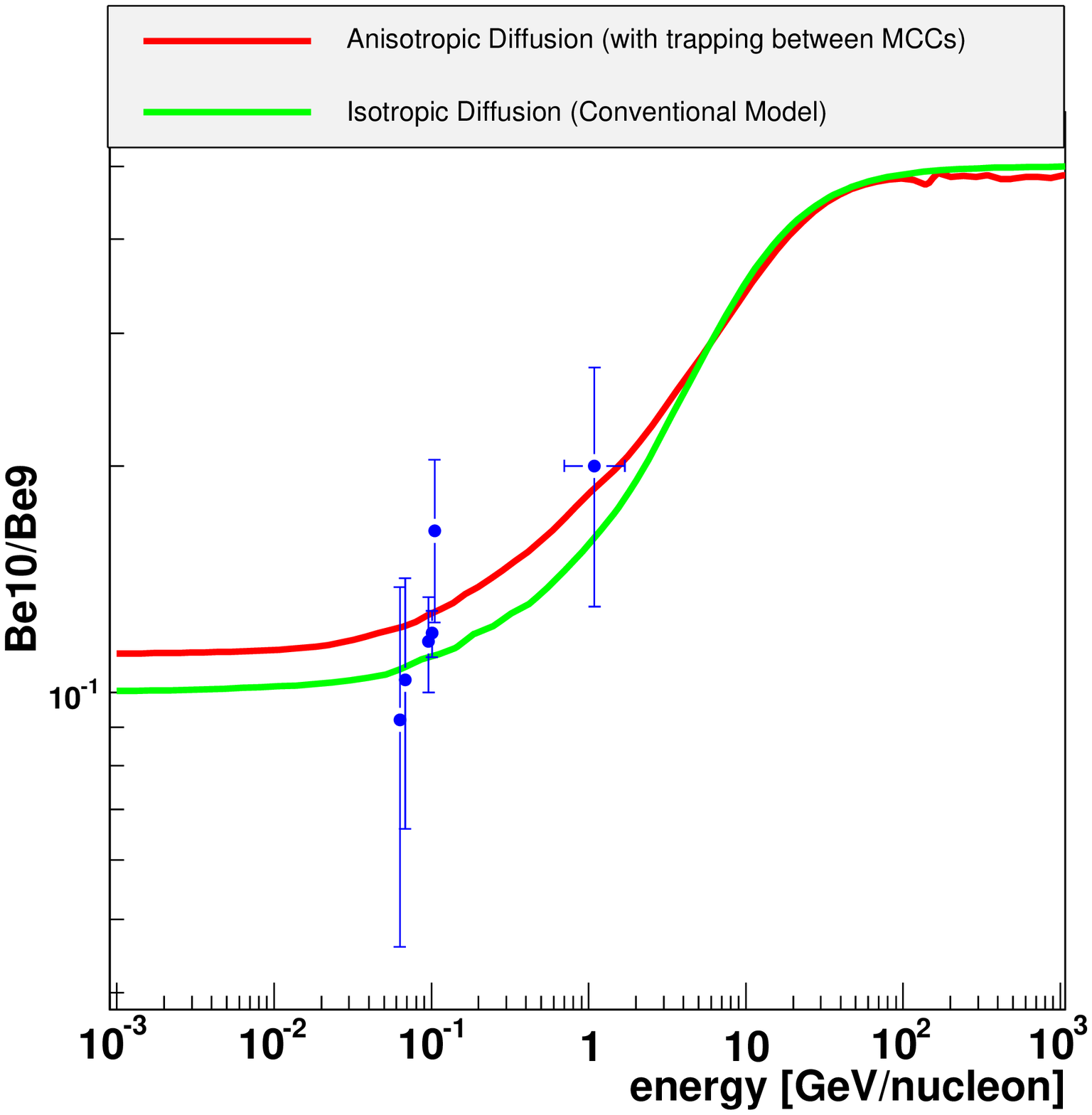}
 \caption[]{Comparison of the secondary/primary ($B/C$) ratio and unstable/stable
 ($^{10}Be/^9Be$) ratio in the conventional model with isotropic propagation
 and  model with anisotropic propagation.}
 \label{fig6}
\end{center}
\end{figure}

The transport from the disk to the halo is quite uncertain. It should be noted that
the average scale height of SNIa is expected to be about 300 pc (thick disk) and
the ejecta connect to the halo in chimney like structures (see e.g. Ref.
\cite{breitschwerdt,breitschwerdt1} and references therein), which can drive
magnetic field lines towards high altitudes ( $\approx$ 1 kpc), thus facilitating
the transport to the halo by the fast parallel diffusion. This was simulated as an
enhanced convection term starting at $v_0=250 km/s$ at 100 pc above the disk and
then increasing with the distance $z$ above the disk as $dv/dz=37 km/s$. The
remaining GALPROP parameters can be found in Ref. \cite{gebauer}. As shown in Figs.
\ref{fig4} and \ref{fig6} the  antiproton flux, the B/C ratio and the
$^{10}Be/^9Be$ ratio are all well described by this set of parameters. Clearly, the
excess of diffuse gamma rays can be well described together with the B/C ratio and
the antiprotons.

In summary, we do not believe that  the DMA interpretation of the EGRET is
"excluded by a large margin" because of the overproduction of antiprotons as
claimed by Bergstr\"om et al. \cite{bergstrom1}. Such a statement  is only valid
{\it within their propagation model based on isotropic propagation}. Anisotropic
models with different propagation in the halo and the disk can perfectly describe
all observations inclu\-ding DMA.

\section{Positron Annihilation in our Galaxy}\label{integral}

Additional support for the propagation picture with anisotropic propagation, as
discussed in  section \ref{prop2}, comes from the positron annihilation signal
observed by the INTEGRAL satellite \cite{integral,integral1}. Since positron
annihilation is only efficient at non-relativistic energies, the positrons must
have energies in the MeV range. Sources of such positrons are largely coming from
the decay of radioactive nuclei expelled by dying stars, especially SNIa, since in
this case the core makes up a large fraction of the mass. This makes it easier for
the positrons to escape from the relatively thin layer of the ejecta. Light curves,
which are sustained first by the gamma rays in the shock waves and later by the
electrons and positrons, suggest that only a few percent of the positrons escape
from the ejecta and can annihilate outside after thermalization. Positrons
annihilating inside the ejecta will produce also gamma rays, but these will not be
visible as a single  511 keV line because of further interactions in the shock
wave.

The main observation is that the 511 keV  line from the annihilation between
thermalized positrons and electrons (either free or bound in nuclei) is largely
confined to the bulge with a bulge/disk (B/D) ratio of a few
\cite{integral,integral1}, although additional data suggests a lower ratio
\cite{integral2}. Taking the dominant source to be SNIa, one would expect a B/D
ratio to be well below one, because of the higher mass in the disk and the higher
rate of SNIa explosions expected in the thick disk as compared to the bulge
\cite{prantzos}. An additional problem presents the observation of the 1.8 MeV line
from the $^{26}Al$ radioactive isotope, which has clearly been observed, both in
the bulge {\it and} the disk by the Comptel detector on NASA's CGRO observatory
\cite{al26}. These nuclei are thought to be produced by nucleosynthesis in massive
stars in the thin disk and yield in their decay on average 0.85 positrons. Because
of their non-relativistic speed and high charge $^{26}Al$ nuclei loose their kinetic energy
rapidly and decay with a half life time of about $10^6$ years close to the position
where they were created. The observed flux of positron annihilation in the disk
seems to be saturated already by the positrons from $^{26}Al$ and $^{44}Ti$ decays.\cite{integral}
 So what happened to the positrons from SNIa explosions in the
disk?

In an anisotropic propagation model a large B/D ratio for the positrons is
expected, since the bulge is a much more extended object than the disk, so the
particles have much more time to thermalize and annihilate in the bulge than in the
disk before reaching the halo. Once in the halo they move away from the disk, where
there is hardly any gas for them to annihilate. As mentioned before the fountain
like structures of SNR (see e.g. Refs. \cite{breitschwerdt,breitschwerdt1} and
references therein) can drive magnetic field lines towards high altitudes ($\approx$ 1 kpc),
thus facilitating the transport to the halo by the fast parallel
diffusion for the relativistic positrons.

 A fraction of the positrons escaping the disk may
move by the poloidal field to the bulge, thus enhancing the B/D ratio
\cite{prantzos}. So the problem of the large B/D ratio for positron annihilation is
intimately related to the propagation of the positrons.  In a conventional
propagation model without magnetic fields and homogeneous gas distributions the
positrons annihilate near their source \cite{jean1,jean} and one must resort to new
positron sources specific for the bulge, like DMA of very light WIMPS
\cite{boehm,boehm1}. The WIMP masses have to be below of few MeV, since
else they would be visible by synchrotron radiation above the limits set by Comptel
and EGRET.\cite{beacom,boehm2} In anisotropic propagation models the low strength
of the annihilation signal in the disk is simply a consequence of the fact that
the disk is so thin, so positrons can escape easily to the halo, where they find no
partners to annihilate. So it is the same solution required by the ratio of antiprotons
and EGRET excess of gamma rays: fast diffusion perpendicular to the disk either
by the regular magnetic fields or convection.
 In this case no unnatural light WIMPs
are  needed. Such light WIMPs would need additionally new gauge bosons to
have a large enough annihilation cross section compatible with Eq. \ref{eq1}.
\section{Constraints from direct dark matter search experiments}\label{direct}
\begin{figure}
\begin{center}
 \includegraphics [width=0.65\textwidth,clip]{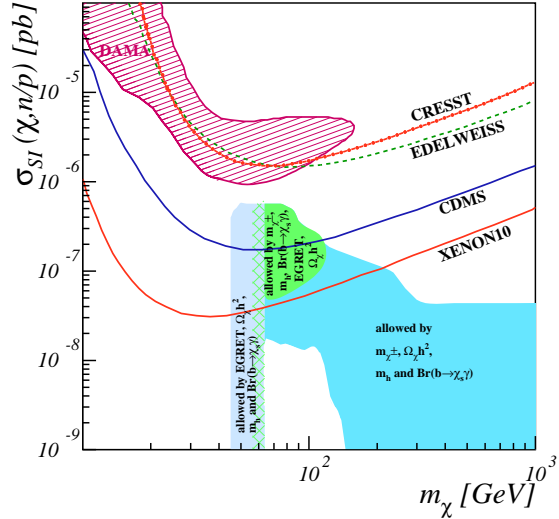}
   \caption[]{The spin-independent cross section for neutralino-nucleon scattering
   compared with experimental constraints: the region above the XENON10 limit is excluded.
   Details can be found in Ref. \cite{deger}. }
 \label{fig7}
\end{center}
\end{figure}
Direct detection experiments search for signals of dark matter particles
elastically scattering off the nuclei in their detector. The event rate is given by
$\Gamma=<\sigma v> n$, where $n$ is the local number density of WIMPS, $v$ the
velocity between detector and WIMP and $\sigma$ the scattering cross section. The
brackets indicate that the ave\-rage over the velocity distribution and
corresponding cross sections has to be taken. No positive results have been
reported and experimental limits on the cross section have been derived under the
assumption that the local density of WIMPS corresponds to $0.3~GeV/cm^3$, as
estimated from the rotation curve (see e.g. a recent review by
Spooner\cite{spooner}). The best limit obtained by the XENON10 experiment is about
$5.10^{-8}$ pb at 90\% C.L. for a 50 GeV WIMP.\cite{xenon10}  This is below the
cross section limit expected from the EGRET data, if one assumes the minimal
supersymmetric model to be valid. The region allowed by all constraints (WMAP, LEP
limits on charginos and Higgses, $b\rightarrow s\gamma$ branching ratio, EGRET mass
limits) is shown in Fig. \ref{fig7} together with the limits from some typical
direct dark matter search experiments.\cite{deger} No signal was found sofar, so
the regions above these lines are excluded. Naively one could conclude that the DMA
interpretation of the EGRET excess combined with all electroweak and relic density
constraints (green region in Fig. \ref{fig7})
 is excluded by the XENON10 data. However, there are severe caveats. First of all, the event
rate in these experiments is proportional to the cross section {\it and} the local
WIMP density. Both have large uncertainties: the cross section is model dependent
and has only been calculated in the minimal supersymmetric models. Although the EGRET data is
perfectly consistent with supersymmetry, it does not prove supersymmetry and in e.g.
extra dimension models the cross sections could be very different. Secondly the local
relic density has large uncertainties, because the densities obtained from the rotation
curve do not say anything about the clustering of dark matter, but yield
only an averaged density. From N-body simulations one knows that dark matter is
clumpy, as can be expected already from the fact that galaxies are formed by
dark matter starting to pull together from the initial density
fluctuations in the early universe. The number of
clumps is very large because of the steep decrease of the mass spectrum with
increasing mass ($\propto M^{-2}$) and the lightest clumps are very light (about
$10^{-6}M_\odot$, see Ref.\cite{moore}). But even with this large number of clumps the
probability of finding a clump in the solar neighbourhood  is small, so the direct
searches are more likely to observe first interactions from the diffuse component,
which originates from the tidal stripping of the clumps. How many are disrupted is
an ongoing debate (see e.g. Refs. \cite{moore1,dokuchaev}), but it certainly depends
sensitively on the dark matter profile of the clumps. These profiles cannot be
calculated yet reliably from N-body simulations. It is concei\-vable that the density
of the diffuse component is an order of magnitude less than the clumpy component,
so any statements concerning exclusions from direct dark matter detection
experiments should take this into account.

\section{Conclusion}\label{conclusion}
 With an anisotropic propagation model the amount of
antiprotons expected from DMA annihilation can be reduced by one to two orders of
magnitude. Therefore the claim by Bergstr\"om et al.\cite{bergstrom1} that the DMA interpretation of
the EGRET excess of diffuse Galactic gamma rays is excluded ``by a large margin''
is strongly pro\-pagation model dependent. It only applies for their simplified
propagation model, which assumes the same diffusion in the halo and in the disk. An
anisotropic propagation model with different propagation in the halo and the disk
can reconcile the EGRET excess with the antiproton flux and the ratios of
secondary/primary and unstable/stable nuclei. In addition the difference in
geometry between bulge and disk leads to much more radiation and annihilation of
positrons and electrons in the bulge as compared to the disk, thus alleviating the
need to introduce new sources of positrons and electrons for the bulge to explain
the INTEGRAL excess of the 511 keV line in the bulge\cite{boehm}.
Direct dark matter search experiments reach the cross section range expected from the
EGRET excess.
However, the experimental limits are inversely proportional to these cross sections times
the local relic density. Both are uncertain: for the cross section  one assumes the
WIMPS are the lightest supersymmetric partners
of the minimal supersymmetric model with gravity inspired breaking of supersymmetry, while
for the local relic density one assumes the WIMPS are smoothly distributed instead of
the expected distribution in clumps. The clumpy nature can drastically reduce the
local density if we are not located inside a clump.

In summary we consider DMA is a viable explanation of the EGRET excess of diffuse
Galactic gamma rays, especially since it is observed with the same shape of the
fragmentation of mono-energetic quarks in all sky directions and the intensity
distribution of the excess traces the DM profile, as shown independently by the
rotation curve, the gas flaring and the N-body simulation of the disruption of the
Canis-Major satellite galaxy.
\section{Acknowledgements} I wish to thank the organizers for the kind invitation
to this splendid conference. Furthermore I thank P. Blasi, D. Breitschwerdt and N.
Prantzos for helpful discussions and my close collaborators I. Gebauer, D. Kazakov,
M. Weber and V. Zhukov for their contributions.

 This work was supported by the BMBF (Bundesministerium f\"ur Bildung und Forschung) via the DLR
(Deutsches Zentrum f\"ur Luft- und Raumfahrt).
\bibliographystyle{aa}

\end{document}